\documentclass[%
 reprint,
 amsmath,
 amssymb,
onecolumn,
 aps,
 prx,
floatfix,
longbibliography,
superscriptaddress
]{revtex4-2}

\makeatletter
\def\@bibdataout@aps{%
 \immediate\write\@bibdataout{%
  @CONTROL{%
   apsrev41Control,author="08",editor="1",pages="0",title="0",year="1"%
  }%
 }%
 \if@filesw
  \immediate\write\@auxout{\string\citation{apsrev41Control}}%
 \fi
}%
\makeatother % Phew. 

\usepackage{graphicx}% Include figure files
\usepackage{dcolumn}% Align table columns on decimal point
\usepackage{bm}% bold math
\usepackage{layouts}
\usepackage[colorlinks,linkcolor=blue,anchorcolor=blue,citecolor=blue,urlcolor=blue]{hyperref}% add hypertext capabilities
\usepackage{float}
\usepackage{makecell}% for table
\usepackage{verbatim}
\usepackage{color}
\usepackage{lineno}% Enable numbering of text and display math
\usepackage[thicklines]{cancel}
\usepackage{ulem}
\usepackage{cancel}
\usepackage{bm}
\usepackage{multirow}
\usepackage{xcolor}
\usepackage{graphicx}
\usepackage{cancel}
\usepackage{amsthm}
\usepackage{mathtools}
\usepackage{physics}
\usepackage{ulem}
\DeclarePairedDelimiterX\braketOP[3]{\langle}{\rangle}{#1\,\delimsize\vert\,\mathopen{}#2\,\delimsize\vert\,\mathopen{}#3}%
\DeclarePairedDelimiterX\braketReOP[3]{\langle}{\rangle}{#1\,\delimsize\Vert\,\mathopen{}#2\,\delimsize\Vert\,\mathopen{}#3}%
\DeclarePairedDelimiterX\outerp[2]{\lvert}{\rvert}{#1\delimsize\rangle\!\delimsize\langle#2}%
\DeclarePairedDelimiterX\projector[1]{\lvert}{\rvert}{#1\delimsize\rangle\!\delimsize\langle#1}%

\begin{document}

%/\linenumbers

\preprint{APS/123-QED}

\title{Observation of the electric Breit-Rabi Effect}% Force line breaks with% Force line breaks with \\
%\thanks{These authors contributed equally to this work.}%

\affiliation{Hefei National Research Center for Physical Sciences at the Microscale, School of Physical Sciences, University of Science and Technology of China, Hefei 230026, China}
\affiliation{Hefei National Laboratory, University of Science and Technology of China, Hefei 230088, China}

\author{S.-Z. Wang}%
\thanks{These authors contributed equally to this work.}
\affiliation{Hefei National Research Center for Physical Sciences at the Microscale, School of Physical Sciences, University of Science and Technology of China, Hefei 230026, China}

\author{S.-B. Wang}%
\thanks{These authors contributed equally to this work.}
\affiliation{Hefei National Research Center for Physical Sciences at the Microscale, School of Physical Sciences, University of Science and Technology of China, Hefei 230026, China}

\author{Z.-J. Tao}%
\affiliation{Hefei National Research Center for Physical Sciences at the Microscale, School of Physical Sciences, University of Science and Technology of China, Hefei 230026, China}

\author{T. Xia}%
\thanks{Corresponding author: txia@hfnl.cn}
\affiliation{Hefei National Laboratory, University of Science and Technology of China, Hefei 230088, China}

\author{Z.-T. Lu}%
\thanks{Corresponding author: ztlu@ustc.edu.cn}
\affiliation{Hefei National Research Center for Physical Sciences at the Microscale, School of Physical Sciences, University of Science and Technology of China, Hefei 230026, China}
\affiliation{Hefei National Laboratory, University of Science and Technology of China, Hefei 230088, China}

\date{\today}% It is always \today, today,

\begin{abstract}
The response of an atom to external electric and magnetic fields can reveal fundamental atomic properties. It has long been verified that, in a static magnetic field, those atomic energy levels with hyperfine interactions shift according to the Breit-Rabi formula, which introduces nonlinear dependence on the magnetic field. On the other hand, the corresponding Breit-Rabi dependence on a static electric field has not been observed before due to a combination of experimental challenges. Here we precisely measure the Stark shift of the $6s^2\ ^1S_0\ \leftrightarrow\ 6s6p\ ^1P_1$ transition of $^{171}$Yb ($I$ = 1/2) with cold atoms held by an optical dipole trap in a static electric field up to 120 kV/cm. We observe the electric Breit-Rabi effect displaying high-order ($E^4$ and $E^6$) DC Stark shifts. These effects arise from the influence of the strong electric field on hyperfine interactions.

\end{abstract}

\maketitle

{\center\section{Introduction}}
Extensive research has been conducted on the atomic Zeeman and Stark effects since both effects were discovered over a century ago \cite{ZEEMAN1897,STARK1913}. These effects are sensitive probes of fundamental atomic structure properties. They are also useful, and, indeed, are widely applied to control energy level shifts. For an atomic level with hyperfine interaction, its shift under the influence of an external DC magnetic field is described by the Breit-Rabi formula  \cite{BreitRabi1931}, which leads to approximately linear dependences on the magnetic field in both the weak-field Zeeman effect and the strong-field Paschen-Back effect \cite{Paschen1912}. In the intermediate-field region, where the energy shift is comparable to the hyperfine splitting, the dependence clearly deviates from linearity \cite{NZL2018,NZL2019,SuppressionNZL2018}. These dependencies have been accurately measured, and the Breit-Rabi formula confirmed, over a wide range of magnetic fields. \par

For an atomic level with hyperfine interaction, its Stark shift is expected to exhibit an electric-field version of the Breit-Rabi effect. In the intermediate-field region, the energy shift is predicted to deviate from its usual proportionality to the square of the electric field ($E^2$). The electric Breit-Rabi effect has appeared in textbooks \cite{Steck2019} and theoretical calculations \cite{hyperfinestarkAngel1968,hyperfinestarkSchmieder1972}, and has been applied to determine level-crossing points, which in turn led to hyperfine structure constants of high-lying excited states \cite{PhysRevA.75.022502,Auzinsh2006}. However, the electric Breit-Rabi effect has never been verified with experimental measurements due to a combination of technical challenges. In this work, by precisely measuring the DC Stark shift of a transition in the ytterbium atom, we observe the electric Breit-Rabi effect displaying high-order ($E^4$ and $E^6$) dependences, thus establishing experimental validation of a fundamental atomic effect. Atoms in a laser field also exhibit Stark effects beyond the $E^2$ dependence \cite{PhysRevLett.119.253001,PhysRevA.94.042308}, known as hyperpolarizability, which arises from the 4$^{\rm th}$-order perturbation in the calculation without altering hyperfine interaction. In contrast, the electric Breit-Rabi effect originates from the coupling between the traditional Stark effect and hyperfine interaction. As shown in this work, at the 2$^{\rm nd}$-order perturbation level, the effect already exhibits high-order dependencies ($E^4$ and $E^6$).\par

The neutral ytterbium (Yb) atom possesses a rich Group-II-like energy level structure, providing long-lived clock states, long coherence times for the nuclear spin states, and transitions suitable for laser cooling and trapping. These characteristics combine to make Yb a favorite choice for a wide range of research, including optical clocks \cite{opticalclock2013}, quantum computation \cite{quantuminformation2009,quantuminformation2022,quantuminformation2023}, Bose-Einstein condensation and Fermionic degenerate gases \cite{quantumgas2007,quantumgas2013}. The Stark effect often plays a crucial role in these works, for example, in evaluating the systematic error of an atomic clock due to blackbody radiation \cite{opticalclock2014,BBR2024}, in realizing electric Feshbach resonances \cite{electricFeshbach1998,electricFeshbach2007}, and in spin-state manipulation \cite{spinmanipulation2024}.\par

In this work, we measure the Stark shift of the $6s^2\ ^1S_0\ \leftrightarrow\ 6s6p\ ^1P_1$ transition in $^{171}$Yb (nuclear spin $I$ = 1/2) atoms held by an optical dipole trap. The $6s6p\ ^1P_1$ energy level has a relatively small hyperfine splitting of 318 MHz \cite{hyperfine2016}. As a result, when the electric field is in the practically reachable range of 30 - 100 kV/cm, the Stark interaction of this level is said to enter the intermediate-field region as its tensor energy shift, on the order of 20 - 200 MHz, becomes comparable to the hyperfine splitting. Using cold atoms for this measurement offers multiple advantages: the sample is of pure $^{171}$Yb isotope; the sample fits through a small gap between electrodes so that a high electric field can be applied; Doppler broadening is reduced, and systematic errors induced by quantum interference suppressed. Together, these advantages lead to improved measurement precisions.\par

\vspace{\baselineskip}
{\section{Electric Breit-Rabi formula}}
The Hamiltonian for Stark interaction can be expressed in the $I$-$J$ coupled basis as \cite{hyperfinestarkSchmieder1972,Steck2019}
\begin{equation}\label{equ:StarknonlinearV2}
\begin{aligned}
\langle F'm_F'|H_{\rm S}|F''m_F''\rangle = -\frac{1}{2} \left( \alpha_{s} \mathcal{I} + \alpha_{t} \mathcal{Q} \right) E^2,
\end{aligned}
\end{equation}
where $E$ is the static electric field, its direction defined as the quantization axis. The scalar shift led by the static scalar polarizability ($\alpha_s$) is the same for all hyperfine states, and $\mathcal{I}=\delta_{F'F''} \delta_{m_F'm_F''}$. On the other hand, the tensor shift (with $\alpha_t$) varies due to hyperfine mixing and is described by the matrix element 
\begin{equation}\label{QFMV2}
\begin{aligned}
&\mathcal{Q} =  \sqrt{\frac{(J'+1)(2J'+1)(2J'+3)(2F'+1)(2F''+1)}{J'(2J'-1)}} \\
&\times(-1)^{I+J'-m_F'} \begin{Bmatrix} F' & 2 & F'' \\ J' & I & J' \end{Bmatrix} \begin{pmatrix} F' & 2 & F'' \\ m_F' & 0 & -m_F'' \end{pmatrix} \delta_{m_F'm_F''}.
\end{aligned}
\end{equation}\par

By diagonalizing the Hamiltonian ${H=A_{\rm hfs}\boldsymbol{I}\cdot\boldsymbol{J}+H_{\rm S}}$ , including both the hyperfine and Stark interaction, we obtain differential DC Stark shifts for the $6s^2\ ^1S_0\ |Fm_F\rangle\ \leftrightarrow\ 6s6p\ ^1P_1\ |F'm_F'\rangle\ $ transitions in $^{171}$Yb:
\begin{equation}\label{equ:k1712}
\begin{aligned}
\Delta \nu=K_{F',m_F'}E^2+f_{F',m_F'}(E),
\end{aligned}
\end{equation}
where $K_{F',m_F'}$ represents the differential DC Stark shift rate (see SI Appendix). The first term in Eq. \ref{equ:k1712} describes the commonly known 2$^{\rm nd}$-order DC Stark shift, which arises from the mixing between energy levels of opposite parities, e.g. between $S$ and $P$ levels. The second term describes high-order Stark shifts that arise from state mixing between different hyperfine levels due to the action of the electric field on $I$-$J$ coupling, and these mixings form superpositions of states with the same parity.\par
\begin{figure}[htbp]
	\centering
	\includegraphics[width=10.5cm]{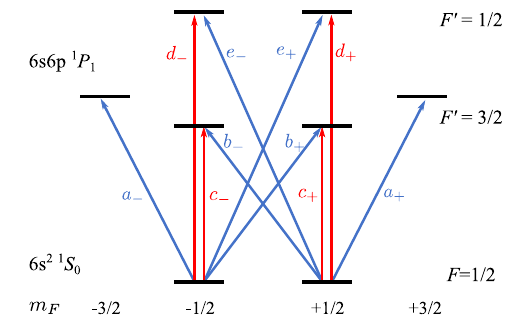}
	\caption{Energy levels and transitions of $^{171}$Yb. The $6s^2\ ^1S_0\ \leftrightarrow\ 6s6p\ ^1P_1$  transition has a wavelength of 399 nm and a natural linewidth $\Gamma=2\pi\times28.9$ MHz \cite{naturelinewidth2004}. The hyperfine splitting between $F'=1/2$ and $F'=3/2$ of $6s6p\ ^1P_1$ is 318.5 MHz. $\pi$ transitions (red) are labeled $c_{\pm}$ and $d_{\pm}$; $\sigma$ transitions (blue) labeled $a_{\pm}$, $b_{\pm}$ and $e_{\pm}$. 
}
	\label{fig:Figure(1)}
\end{figure} 
For the case of the $E$ field pointing in the quantization axis, state mixing due to the tensor Stark interaction only occurs between states of the same $m_F$. In the $6s6p\ ^1P_1$ level of $^{171}$Yb (Fig. \ref{fig:Figure(1)}), the high-order terms $f_{3/2,\pm3/2}(E)$ equal to zero because there are no state mixing for the $^1P_1\ |3/2,\pm3/2\rangle$ states. On the other hand, the hyperfine states of $m_F=\pm1/2$ experience state mixing under the tensor Stark effect, and the new energy eigenstates are superpositions of $^1P_1\ |3/2,1/2\rangle$ and $^1P_1\ |1/2,1/2\rangle$ (or $^1P_1\ |3/2,-1/2\rangle$ and $^1P_1\ |1/2,-1/2\rangle$). We derive an electric-field version of the Breit-Rabi formula (see SI Appendix), 
\begin{equation}\label{equ:DC stark 171}
\begin{aligned}
f_{F',m_F'}(E)&=
        \pm\frac{1}{4}\left(\alpha_tE^2+3A_{\rm hfs}-3A_{\rm hfs}\sqrt{1
    +\frac{2}{3}x+x^2}\right)\\
    &=\pm\frac{A_{\rm hfs}}{3}\left(-\frac{\alpha_t^2}{A_{\rm hfs}^2}E^4+\frac{\alpha_t^3}{3A_{\rm hfs}^3}E^6+\mathcal{O}(\frac{\alpha_t^4}{A_{\rm hfs}^4}E^8)\right),  
\end{aligned}
\end{equation}
where the symbol $\pm$ correspond to the energy shifts of the upper and lower states. In analogy to the magnetic-field version, the unitless parameter $x=\alpha_tE^2/A_{\rm hfs}$ describes the size of the field effect relative to that of the hyperfine effect. In the weak-field case ($x \ll 1$), it is well known that the Stark shift is proportional to $E^2$. However, in the intermediate-field region ($x \sim 1$), the term $f_{F',m_F'}(E)$ displays high-order Stark effects. In this study, we explore the region of $0 < x < 0.9$.\par
%\vspace{\baselineskip}
{\section{Experimental setup}}

\begin{figure}[htbp]
	\centering
	\includegraphics[width=12.5cm]{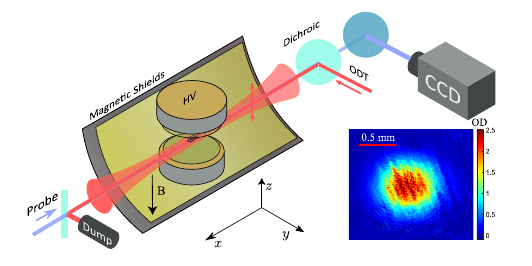}
	\caption{Experimental setup. $1\times10^6$ $^{171}$Yb atoms are trapped in the optical dipole trap (ODT). A pair of copper electrodes generate a static $E$ field up to 120 kV/cm along $z$. A cos($\theta$) coil inside the magnetic shields provides a bias $B$ field of 20 mG, also along $z$. The probe laser beam propagates along $x$. The absorption image is captured using a CMOS camera, and the colorbar represents the optical density (OD). 
The figure is created using the open-source ComponentLibrary\cite{ComponentLibrary}.
}
	\label{fig:Figure(2)}
\end{figure}

$^{171}$Yb atoms are captured and cooled by a two-stage magneto-optical trap system. The cold atoms are transported to a neighboring science chamber with an optical dipole trap (ODT) for spectral measurement. Sample preparation, including atom trapping and transport, as well as high-voltage ramping, takes a total of approximately 20 s. As shown in Fig. \ref{fig:Figure(2)}, the ODT propagates along the $x$ direction, and is linearly polarized along the $z$ axis. Its wavelength of 1035.8 nm is set at the magic wavelength of the $6s^2\ ^1S_0\ \leftrightarrow\ 6s6p\ ^3P_1$ transition \cite{magicwavelength2020}. With a power of 30 W, radius of 60 $\mu$m, and Rayleigh length of 7 mm, it achieves a trap depth of 100 $\mu$K. In the ODT, more than $1\times10^6$ atoms at 40 $\mu$K form a cloud of 100 $\mu$m size transversely and 8 mm longitudinally. At a vacuum pressure of approximately $1\times10^{-11}$ torr, the trap lifetime of atoms in the ODT, positioned between the electrodes, is measured to exceed 70 s.  \par

The atoms are positioned at the center between the electrodes. A uniform electric field is generated by a pair of parallel (tilt $<$ 2 mrad) copper electrodes with a diameter of 16 mm and a separation distance measured optically to be 1.08(5) mm, which directly leads to a conversion factor between the voltage and the electric field to be 9.26(43) (kV/cm)/kV. Meanwhile, Stark shift measurements are performed to calibrate the same conversion factor, resulting in a more precise value of 8.654(14) (kV/cm)/kV. With a difference of 1.4 times the standard deviation, the two values are statistically consistent with each other, and are combined to produce a weighted average value of 8.655(14) (kV/cm)/kV. For this calibration, we measure the Stark shifts of the $6s^2\ ^1S_0\ \leftrightarrow\ 6s6p\ ^3P_1$ transition of the $^{176}$Yb isotope (nuclear spin $I = 0$, no hyperfine structure), and compare results with the published values from Li $et\ al$. \cite{3P1stark1995} (see SI Appendix). In the work of Li $et\ al$., measurements were carried out with an atomic beam through a pair of electrodes with a large spacing of 1.0163(3) cm , which was measured using machinist’s gauge blocks \cite{3P1stark1995}. \par

A probe laser at the wavelength of 399 nm is used to perform spectroscopy on the $6s^2\ ^1S_0\ \leftrightarrow\ 6s6p\ ^1P_1$ transition and measure its Stark shifts. Combined with the ODT, the probe laser beam also propagates along the $x$-direction (Fig. \ref{fig:Figure(2)}). The beam diameter of 2 cm is much larger than the size of the atom cloud so that its intensity on the atoms is approximately uniform at 50 $\mu$W/cm$^2$ (to be compared to the saturation intensity of 58 mW/cm$^2$). A sideband generated by a fiber-EOM is frequency locked to a ULE cavity, and the carrier is used to scan across spectra as the EOM frequency is changed.\par

Fig. \ref{fig:Figure(1)} illustrates the hyperfine structure of the $6s^2\ ^1S_0\ \leftrightarrow\ 6s6p\ ^1P_1$ transitions in $^{171}$Yb, with each transition labelled by a letter from $a_{\pm}$ to $e_{\pm}$. For certain measurements, the detection beam is set to be linearly polarized along $z$ to induce $\pi$ transitions labelled by $c_{\pm}$ and $d_{\pm}$; for other measurements, linear polarization along $y$ is used to induce $\sigma$ transitions labelled by $a_{\pm}$, $b_{\pm}$ and $e_{\pm}$. The absorption signal is captured by a CMOS camera with an exposure time of 0.4 ms as shown in Fig. \ref{fig:Figure(2)}. The ODT is turned off at 2 ms before each camera exposure in order to avoid light shifts.\par
%\vspace{\baselineskip}
{\section{Spectroscopy}}

The interplay between hyperfine and Stark interactions not only induces energy shifts described by the electric Breit-Rabi formula, but also causes transition rates to vary. As the electric field increases, the eigenstates evolve from the pure $I$-$J$ coupling states $|Fm_F\rangle$ to their superpositions $\overline{|Fm_F\rangle}$, leading to changes in transition rates. In particular, transitions $b_{\pm}$ become more and more suppressed as the electric field increases. More details of the transition-rate measurements are provided in SI Appendix.\par
\begin{figure}[htbp]
	\centering
	\includegraphics[width=10.5cm]{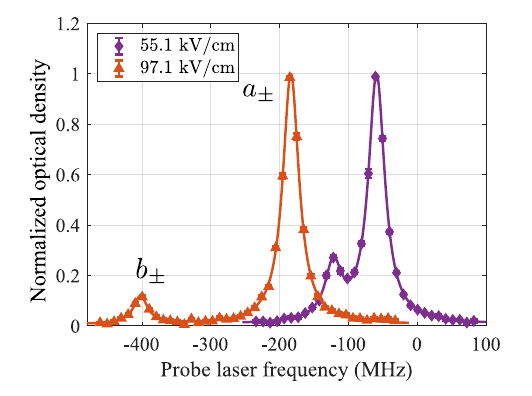}
	\caption{ The spectra of transitions $a_{\pm}$ (higher peaks) and $b_{\pm}$ (lower peaks) under the $E$ field of 55.1 kV/cm (puple) and 97.1 kV/cm (red), respectively. The vertical axis represents the normalized optical depth, with each spectrum normalized independently.
}
	\label{fig:Figure(3)}
\end{figure}
Each absorption spectrum displayed in Fig. \ref{fig:Figure(3)} contain two peaks: the higher peak on the right is due to transitions $a_{\pm}$, $^1S_0\ |1/2,\pm1/2\rangle\ \leftrightarrow\ ^1P_1\ |3/2,\pm3/2\rangle$ , and lower peak on the left is due to transitions $b_{\pm}$, $^1S_0\ |1/2,\pm1/2\rangle\ \leftrightarrow\ ^1P_1\ \overline{|3/2,\mp1/2\rangle}$  . The frequency difference between the two peaks is the tensor Stark splitting, and the ratio between the two peak heights reflects the relative change of these transition rates. \par

At the bias magnetic field of 20 mG, the Zeeman shifts ($\sim$ 30 kHz) are much smaller than the natural linewidth. Moreover, because the Zeeman shifts of the $+ m_F$ and $- m_F$ states are equal but with opposite signs, their combined effect only slightly broadens the spectra lines without shifting the peak positions.\par

%\vspace{\baselineskip}
{\section{High-Order Stark Shifts}}

\begin{figure*}[htbp]
	\centering
	\includegraphics[width=12.7cm]{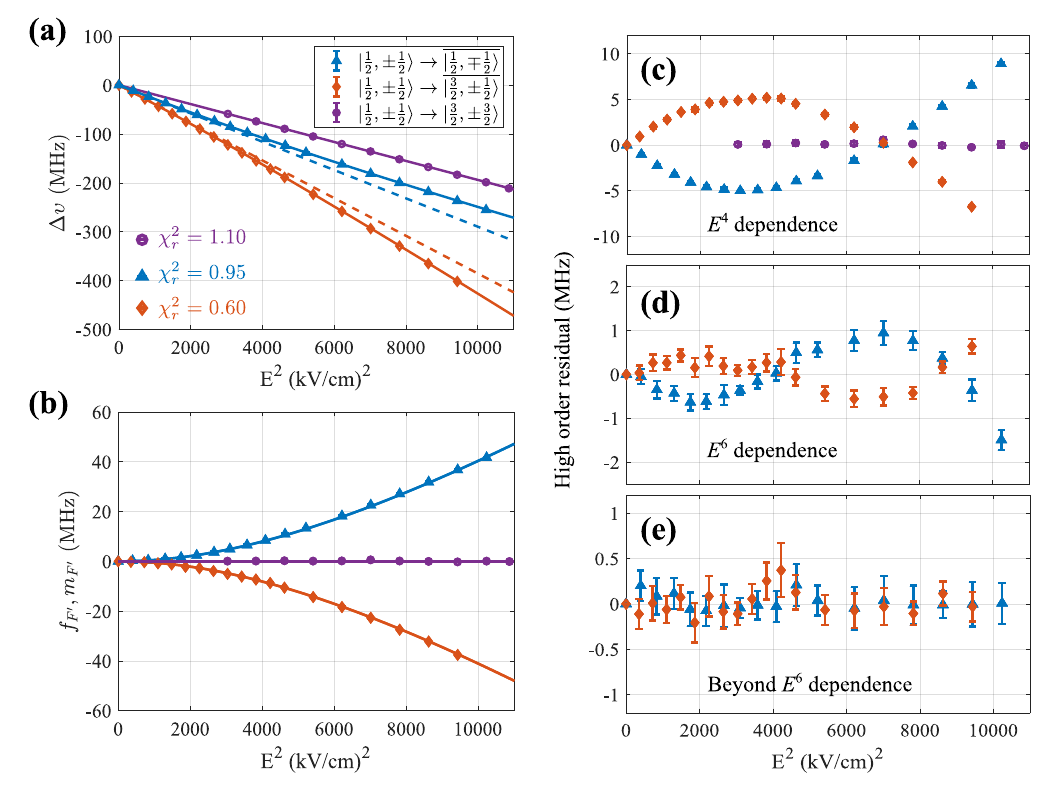}
	\caption{Stark shifts vs. $E^2$. (a) $\Delta\nu$ vs. $E^2$. For clarity, the hyperfine shifts are removed, leaving only the Stark shifts. The error bars are smaller than the symbol size. The solid lines represent fitting by functions $\Delta \nu=K_{F',m_F'}E^2+f_{F',m_F'}(E)$ given in Eq. \ref{equ:k1712}. The red and blue dashed straight lines represent the frequency shifts caused solely by the 2$^{\rm nd}$-order Stark shifts, $K_{F',m_F'}E^2$. (b) $f_{F',m_F'}(E)$ vs. $E^2$. The data points come from the difference between the data in Fig. (a) and the dashed straight lines; The solid lines represent fitting by functions $f_{F',m_F'}(E)$ given in Eq. \ref{equ:DC stark 171}. (c-e) Stark shift data are fit with polynomial functions instead of Eq. \ref{equ:k1712}). (c) Residuals of fitting by the polynomial $\Delta\nu=\beta_2E^2$. (d) Residuals of fitting by the polynomial $\Delta\nu=\beta_2E^2+\beta_4E^4$. (e) Residuals of fitting by the polynomial $\Delta\nu=\beta_2E^2+\beta_4E^4+\beta_6E^6$.
}
	\label{fig:Figure(4)}
\end{figure*}

Fig. \ref{fig:Figure(4)}(a) displays the measured Stark shifts of $6s^2\ ^1S_0\ \leftrightarrow\ 6s6p\ ^1P_1$ in $^{171}$Yb. The purple data points represent shifts of the transitions to the stretched states, $^1S_0\ |1/2,\pm1/2\rangle\ \leftrightarrow\ ^1P_1\ |3/2,\pm3/2\rangle$. Without high-order Stark shifts, their dependence on $E^2$ is linear with a fitted slope of -19.43$\pm0.01_{\rm stat}\pm0.06_{\rm syst}$ kHz/(kV/cm)$^2$. This result is in agreement with, and more accurate by a factor of six than, the value of -19.82 (35) kHz/(kV/cm)$^2$ reported by Kawamura $et$ $al.$ \cite{1P1stark2013}. The red and blue data points represent the Stark shifts of $^1S_0\ |1/2,\pm1/2\rangle\ \leftrightarrow\ ^1P_1\ \overline{|3/2,\pm1/2\rangle}$ and $^1S_0\ |1/2,\pm1/2\rangle\ \leftrightarrow\ ^1P_1\ \overline{|1/2,\mp1/2\rangle}$ transitions, respectively, both experiencing state mixing among hyperfine levels due to Stark interactions. By fitting the data in Fig. \ref{fig:Figure(4)}(a) with the function given in Eq. \ref{equ:k1712} and Eq. \ref{equ:DC stark 171}, the shifts can be attributed to two parts: the regular $2^{\rm nd}$-order and the high-order Stark shifts. The red and blue dashed straight lines represent the frequency shifts caused solely by the $2^{\rm nd}$-order Stark effect ($K_{F',m_F'}E^2$). The difference between the data points and the dashed straight lines, or the solid fit lines and the dashed straight lines, correspond to the high-order Stark shift $f_{F',m_F'}(E)$, and, for clarity, are plotted in Fig. \ref{fig:Figure(4)}(b). As shown clearly in Fig. \ref{fig:Figure(4)}(b). the electric Breit-Rabi formula (Eq. \ref{equ:DC stark 171}, blue and red lines) achieves a good agreement with data ranging from $x=0$ to $x=0.9$ at 100 kV/cm. \par

In order to demonstrate the unambiguous presence of high-order effects, the data in Fig. \ref{fig:Figure(4)}(a) are fitted with $\Delta\nu=\beta_2E^2$. The resulting fit residuals appear as a quadratic polynomial of $E^2$ (Fig. \ref{fig:Figure(4)}(c)), thus demonstrating the presence of an $E^4$ dependence. Likewise, the data in Fig. \ref{fig:Figure(4)}(a) are fitted with $\Delta\nu=\beta_2E^2+\beta_4E^4$ . Its fit residuals form a cubic polynomial of $E^2$ (Fig. \ref{fig:Figure(4)}(d)), which clearly demonstrates the presence of an $E^6$ dependence. Only when data in Fig. \ref{fig:Figure(4)}(a) are fitted with $\Delta\nu=\beta_2E^2+\beta_4E^4+\beta_6E^6$, the fit residue appears flat within error bars (Fig. \ref{fig:Figure(4)}(e)), thus suggesting that $E^8$ and higher -order terms are below the noise background. In conclusion, the electric-field version of the Breit-Rabi formula is verified in this work with experimental measurements.\par

Quantum interference between transitions of different hyperfine levels is commonly present in spectral measurements \cite{Quantuminterference2011,Quantuminterference2013,Quantuminterference2019}. For example, in Kleinert $et$ $al$. \cite{hyperfine2016}, the hyperfine structure constants of the $6s6p\ ^1P_1$ level in $^{171}$Yb were determined using the fluorescence method, where a correction on the order of MHz due to quantum interference was made to obtain the final result of $A_{\rm hfs}$ = -212.3(3) MHz. In this work, we record spectra by the absorption imaging method instead of fluorescence detection. Among other advantages, the absorption method effectively suppresses systematic errors of quantum interference \cite{Quantuminterference2013}. We determine the hyperfine structure constant of the same energy level to be $A_{\rm hfs}$ = -212.4(1) MHz. No correction of quantum interference is needed in this work. \par

Moreover, the high-order Stark shifts induced by the electric Breit-Rabi effect provides additional coefficients that can be used to extract the scalar and tensor polarizability. By fitting the blue and red solid lines in Fig. \ref{fig:Figure(4)}(a) with Eq. \ref{equ:k1712}, and with $A_{\rm hfs}$ given the above value, we obtain both the static tensor polarizability $\alpha_t=-19.35\pm0.08_{\rm stat}\pm0.06_{\rm syst}$ kHz/(kV/cm)$^2$ and the differential static scalar polarizability $\Delta\alpha_s=\alpha_s(^1P_1)-\alpha_s(^1S_0)=57.99\pm0.07_{\rm stat}\pm0.19_{\rm syst}$ kHz/(kV/cm)$^2$ for the $^1P_1$ level of $^{171}$Yb. The values are obtained independently by fitting either the $^1S_0\ |1/2,\pm1/2\rangle\ \leftrightarrow\ ^1P_1\ \overline{|3/2,\pm1/2\rangle}$ or the $^1S_0\ |1/2,\pm1/2\rangle\ \leftrightarrow\ ^1P_1\ \overline{|1/2,\mp1/2\rangle}$ transition, and are consistent with each other, as shown in SI Appendix. They also agree with the values obtained on even isotopes by Kawamura $et$ $al$. \cite{1P1stark2013}, as the polarizability values are shared by all Yb isotopes. \par 
{\section{Conlcusion}}
By measuring the DC Stark shifts of the $6s6p\ ^1P_1$ level in $^{171}$Yb atoms, we have observed the electric Breit-Rabi effect for the first time. The Zeeman effect and the Stark effect are two fundamental effects that describe the interaction between atoms and external fields. While the full Zeeman effect has long been verified experimentally, the measurement of the electric Breit-Rabi effect in this work completes the experimental verification of the Stark effect. \par
The nonlinearity in the magnetic Breit-Rabi effect has been employed to achieve magic conditions at which the differential energy between some hyperfine or Zeeman states becomes insensitive to variation of magnetic field, leading to a longer qubit coherence time in $^{43}$Ca$^+$ \cite{Harty2014} and a frequency standard based on $^{9}$Be$^+$ \cite{Bollinger1985}. Likewise, the high-order part of the electric Breit-Rabi effect offers additional degrees of freedom for tuning atomic level shifts in order to achieve magic conditions for applications in quantum information processing and precision measurements.

\vspace{\baselineskip}
\section{method}

To calibrate the electric field, we measure the DC Stark shift of the $6s^2\ ^1S_0 \leftrightarrow 6s6p\ ^3P_1$ transition (wavelength 556 nm) of the even isotope $^{176}$Yb ($I=0$) and compare it with the differential DC Stark shift rate \( K \) given in Ref. \cite{3P1stark1995}, which is \( K = -15.419(48) \) kHz/(kV/cm)$^2$. The detection light polarization (parallel to the electric field) and the atomic transition in our measurement are the same as those in Ref. \cite{3P1stark1995}. The measurement results are shown in SI Appendix. By linear fitting, we obtain a slope of -1154.7(7) kHz/(kV$^2$). From our measurement and the reference value, we obtain an conversion factor between
the voltage applied and the resulting electric field  of \( 8.654(14) \) (kV/cm)/kV. The error in this result mainly comes from the spectral measurement error in the Ref. \cite{3P1stark1995}.\par
\begin{figure}
\centering
\includegraphics[width=0.9\textwidth]{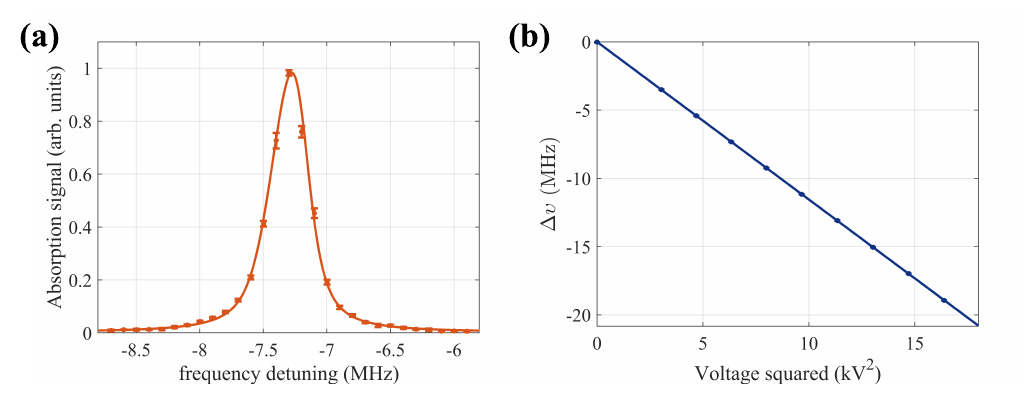}
\caption{Differential DC Stark shift rate of the $6s^2\ ^1S_0|0,0\rangle\leftrightarrow6s6p\ ^3P_1|1,0\rangle$ transition in the even isotope $^{176}$Yb. (a) Spectrum at 2.516 kV, with a resonance frequency of -7.319(12) MHz and a FWHM of 300 kHz. The spectrum is fit with a modified Voigt lineshape, which considers the effect due to acceleration of the atoms by the probe beam. (b) The horizontal axis represents the square of the voltage, and the vertical axis represents the resonance frequency at different voltages. The slope obtained by linear fitting is -1154.7(7) kHz/kV$^2$. The error bars in the figure are from the statistical error of multiple measurements.}
\label{fig:calibration}
\end{figure}
The Breit-Rabi effect is observed by measuring the absorption spectra of the $6s^2\ ^1S_0 \leftrightarrow 6s6p\ ^1P_1$ transition in $^{171}$Yb. The probe laser with wavelength 399 nm can be frequency-scanned using a fiber-EOM. Additional details are described in SI Appendix.\par
Considering both the hyperfine interaction and the DC Stark interaction, we can obtain the energy level shifts of the hyperfine structure under an external electric field as well as the changes in the eigenstate. Additionally, the presence of the static electric field induces state mixing between different hyperfine structures, leading to changes in the transition rates of electric dipole transitions. Additional details are described in the SI Appendix.
%\vspace{\baselineskip}
{\section{Supporting information}}

Considering only the hyperfine interaction, the eigenstate of \( \mathrm{^{171}Yb} \) is denoted as \( \ket{Fm_F} \). The hyperfine interaction Hamiltonian \( H_{\rm hfs} \) is given by
\begin{equation*}\label{equ:hyperfine}
    \bra{F' m_F'}H_{\rm hfs}\ket{F'' m_F''} = \frac{A_{\rm hfs}}{2} \left( F'(F'+1) - I(I+1) - J(J+1) \right) \delta_{F'F''} \delta_{m_F'm_F''},
\end{equation*}
where \( A_{\rm hfs} \) is the magnetic dipole hyperfine constant, \( F'(F'') \) is the total angular momentum quantum number, \( I \) is the nuclear spin, and \( J \) is the total electronic angular momentum. For \( \mathrm{^{171}Yb} \), the nuclear spin \( I=1/2\), hence, the electric quadrupole moment is zero, and there is no electric quadrupole contribution to the hyperfine
interaction.

The DC Stark interaction $H_{\rm S}$ in the presence of hyperfine structure is given by\cite{electricbreitrabi1,electricbreitrabi2,Steck2019}
\begin{equation}\label{equ:Starknonlinear}
\begin{aligned}
\langle F'm_F'|H_{\rm S}|F''m_F''\rangle = -\frac{1}{2} \left( \alpha_{s} \mathcal{I} + \alpha_{t}\mathcal{Q} \right) E^2.
\end{aligned}
\end{equation}
Considering both hyperfine interaction and DC Stark interaction, the total Hamiltonian is given by
\begin{equation}\label{equ:total}
    H_{\rm tot} = H_{\rm hfs} + H_{\rm S}.
\end{equation}\par
The excited state \( 6s6p \, ^1P_1 \) level has two hyperfine structures, labeled as \( F = 3/2\) and \( F = 1/2\), with a total of six distinct Zeeman sublevels. By substituting the energy level structure into Eq.\ref{equ:hyperfine}, Eq.\ref{equ:Starknonlinear} and Eq.\ref{equ:total}, the matrix representation of \( H_{\text{tot}} \) is given by
\begin{equation}\label{Hmatrix}
    \bordermatrix{%
 &|3/2,+3/2\rangle
 & |1/2,1/2\rangle
 & |3/2,1/2\rangle
 &|1/2,-1/2\rangle
 &|3/2,-1/2\rangle
 &|3/2,-3/2\rangle\cr 
 &\frac{A_{\rm hfs}-(\alpha_s+\alpha_t)E^2}{2}  & 0 & 0 & 0& 0&0\cr 
& 0 & -A_{\rm hfs}-\frac{\alpha_sE^2}{2} & \frac{\alpha_tE^2}{\sqrt{2}}& 0 & 0& 0\cr 
& 0 & \frac{\alpha_tE^2}{\sqrt{2}}  &\frac{A_{\rm hfs}-(\alpha_s-\alpha_t)E^2}{2}& 0&0& 0\cr 
& 0 & 0 &0& -A_{\rm hfs}-\frac{\alpha_sE^2}{2} &-\frac{\alpha_tE^2}{\sqrt{2}}& 0\cr 
& 0 & 0 &0& -\frac{\alpha_tE^2}{\sqrt{2}}&\frac{A_{\rm hfs}-(\alpha_s-\alpha_t)E^2}{2} & 0\cr 
& 0 & 0 & 0 &0&0&\frac{A_{\rm hfs}-(\alpha_s+\alpha_t)E^2}{2}\cr 
}
\end{equation}\par

The matrix $H_{\rm tot}$ shows that the matrix contains non-zero off-diagonal elements when both hyperfine interaction and DC Stark interaction are considered. This indicates that the angular momentum eigenstates are mixed, which only occurs between Zeeman sublevels with the same $m_F$ but different $F$. Specifically, mixing occurs between the hyperfine states $|3/2, 1/2\rangle$ and $|1/2, 1/2\rangle$, and between $|3/2, -1/2\rangle$ and $|1/2, -1/2\rangle$. However, in the stretched states where $m_F = \pm 3/2$, there are no pairs of Zeeman sublevels with the same $m_F$, so state mixing does not affect the stretched states. The energy eigenstates of $H_{\rm tot}$ are denoted as $\overline{|F, m_F\rangle}$. In the absence of an electric field, $\overline{|F, m_F\rangle} = |F, m_F\rangle$.

Diagonalizing \( H_{\rm tot} \) yields the atom's eigenenergies and eigenstates under the combined influence of the static electric field and hyperfine interaction. The eigenenergies are given by
\begin{equation}\label{equ:DC stark total}
\begin{aligned}
\mathcal{E}(\overline{|3/2,\pm3/2\rangle})&=\frac{A_{\rm hfs}-(\alpha_s+\alpha_t)E^2}{2}\\
\mathcal{E}(\overline{|3/2,\pm1/2\rangle})&=-\frac{1}{2}\alpha_sE^2+\frac{1}{4}\left(\alpha_tE^2-A_{\rm hfs}+3A_{\rm hfs}\sqrt{1+\frac{2\alpha_tE^2}{3A_{\rm hfs}}+\frac{\alpha_t^2E^4}{A_{\rm hfs}^2}}\right)\\
\mathcal{E}(\overline{|1/2,\pm1/2\rangle})&=-\frac{1}{2}\alpha_sE^2+\frac{1}{4}\left(\alpha_tE^2-A_{\rm hfs}-3A_{\rm hfs}\sqrt{1+\frac{2\alpha_tE^2}{3A_{\rm hfs}}+\frac{\alpha_t^2E^4}{A_{\rm hfs}^2}}\right).
\end{aligned}
\end{equation}
Eq. \ref{equ:DC stark total} represents the energy shifts due to both the hyperfine interaction and the DC Stark effect. After subtracting the energy shifts due to the hyperfine interaction alone, the differential energy shift $\Delta\mathcal{E}_S$ between the excited state $\overline{|F'm_F'\rangle}$ and the ground state $|Fm_F\rangle$ caused solely by the static electric field is given by
\begin{equation}\label{equ:nonlinear}
\begin{aligned}
\Delta\mathcal{E}_S(\overline{|F'm_F'\rangle})&=K_{F',m_F'}E^2+f_{F',m_F'}(E), \\
\end{aligned}
\end{equation}
where the ground state of \(^{171}\text{Yb}\) is \(6s^2 \, ^1S_0\) with \(F = \frac{1}{2}\). This level structure is simple and does not exhibit tensor Stark shifts related to \(F\) and \(m_F\). Therefore, the differential energy shift presented in the equation is solely connected to the hyperfine structure of the excited state. The first term is linearly dependent on the square of the electric field ($E^2$) and arises from the state mixing between the electronic energy levels with opposite parities, e.g. between \( S \) and \( P \) states. The expression for the differential DC Stark shift rate \( K_{F',m_F'} \) is given by
\begin{equation}\label{equ:differential dc Stark shift rate }
\begin{aligned}
K_{F',m_F'} &= -\frac{1}{2}\left(\alpha_s + \frac{3m_F'^2 - F'(F'+1)}{F'(2F'-1)}\alpha_t - \alpha_s^g\right)\quad F' > 1/2 \\
K_{F',m_F'} &= -\frac{1}{2}\left(\alpha_s - \alpha_s^g\right)\quad\quad\quad\quad\quad\quad\quad\quad\quad\quad\quad\quad F' \leq 1/2,
\end{aligned}
\end{equation}
where \( \alpha_s^g \) is the static scalar polarizability of the ground state. The second term in Eq. \ref{equ:nonlinear} is non-linearly dependent on $E^2$ and originates from the state mixing between different hyperfine states with same parity. The non-linear term represents the high-order Stark shifts and, by combining Eq. \ref{equ:nonlinear} and Eq. \ref{equ:differential dc Stark shift rate }, we obtain the electric version of the Breit-Rabi formula
\begin{equation}\label{equ:DC stark 171}
\begin{aligned}
f_{F',m_F'}(E)=
    \begin{array}{lc}
        \pm\frac{1}{4}\left(\alpha_tE^2+3A_{\rm hfs}-3A_{\rm hfs}\sqrt{1
    +\frac{2}{3}x+x^2}\right),
    \end{array}
\end{aligned}
\end{equation}
where $x=\alpha_tE^2/A_{\rm hfs}$, and the symbol $\pm$ correspond to the energy shifts of the upper and lower hyperfine states.\par
By diagonalizing the matrix $H_{\rm tot}$ we obtain not only the energy eigenvalues \( \mathcal{E}(\overline{|F,m_F\rangle}) \) but also the energy eigenstates \( \overline{|F,m_F\rangle} \)
\begin{equation}\label{equ:DC stark 171}
\begin{aligned}
 \overline{\ket{3/2,\pm 3/2}} &= \ket{3/2,\pm 3/2} \\
 \overline{\ket{3/2,\pm 1/2}} &= \sqrt{\frac{r+\left(1+\frac{1}{3}x\right)}{2r}}\ket{3/2,\pm 1/2} \pm {\rm sign}(x) \sqrt{\frac{r-\left(1+\frac{1}{3}x\right)}{2r}}\ket{1/2,\pm 1/2} \\
  \overline{\ket{1/2,\pm 1/2}} &= \sqrt{\frac{r+\left(1+\frac{1}{3}x\right)}{2r}}\ket{1/2,\pm 1/2} \mp {\rm sign}(x) \sqrt{\frac{r-\left(1+\frac{1}{3}x\right)}{2r}}\ket{3/2,\pm 1/2},
\end{aligned}
\end{equation}
where \( r(x) = \sqrt{1+\frac{2}{3}x+x^2} \) and \( \text{sign}(x) \) is the sign function. For \( \mathrm{^{171}Yb} \) in the \( 6s6p \, ^{1}P_1 \) state, \( \alpha_t < 0 \) and \( A_{\rm hfs} < 0 \), hence \( \text{sign}(x) = 1 \). This condition is used throughout this material.
Eq. \ref{equ:DC stark 171} also indicates that in the presence of an electric field, state mixing occurs between Zeeman sublevels with the same \( m_F \) but different \( F \), resulting in \( F \) no longer being a good quantum number. \par
\begin{figure}
\centering
\includegraphics[width=0.9\textwidth]{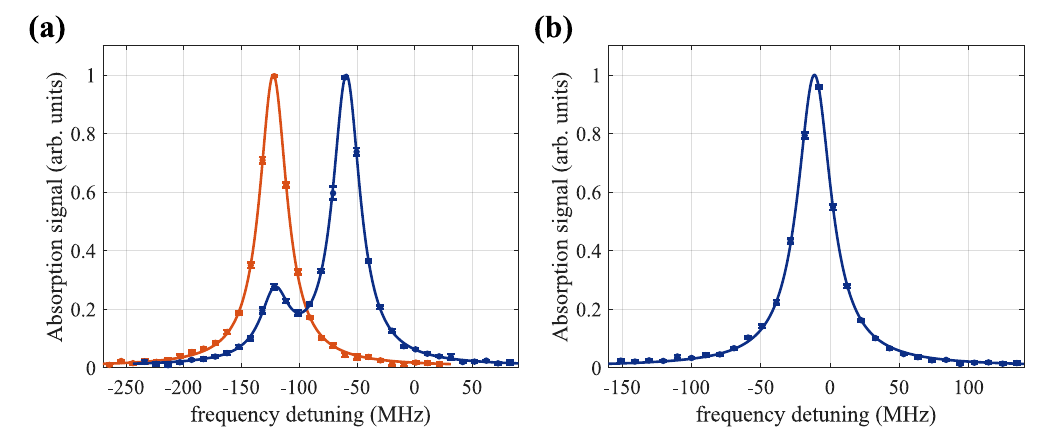}
\caption{Measurement of the $6s^2\ ^1S_0 \rightarrow 6s6p\ ^1P_1$ transition spectrum of $^{171}$Yb. (a) Spectrum of the $^1S_0\ |1/2, m_{F}\rangle \rightarrow {^1}P_1\ \overline{|3/2, m_{F}\rangle}$ transition at an electric field strength of 55.10(9) kV/cm. The orange spectrum is measured with linearly polarized light along the z-axis, corresponding to the $^1S_0\ |1/2, \pm 1/2 \rangle \rightarrow {^1}P_1\ \overline{|3/2, \pm 1/2 \rangle}$ transition. The blue spectrum is measured with linearly polarized light along the y-axis, and the blue solid line is the result of double-peak Lorentzian fitting. The peak on the right corresponds to the $^1S_0\ |1/2, \pm 1/2 \rangle \rightarrow {^1}P_1\ \overline{|3/2, \pm 3/2 \rangle}$ transition. (b) Spectrum of the $^1S_0\ |1/2, \mp 1/2 \rangle \rightarrow {^1}P_1\ \overline{|1/2, \pm 1/2 \rangle}$ transition measured with linearly polarized light along the y-axis at an electric field strength of 19.96(3) kV/cm.}
\label{fig:spectrum} 
\end{figure}
The Breit-Rabi effect is observed by measuring the absorption spectra of the $6s^2\ ^1S_0 \leftrightarrow 6s6p\ ^1P_1$ transition in $^{171}$Yb. The probe laser with wavelength 399 nm can be frequency-scanned using a fiber-EOM. Fig.\ref{fig:spectrum} shows the absorption spectra of three different branches in the $6s^2\ ^1S_0 \leftrightarrow 6s6p\ ^1P_1$ transition described in the main text. The red line in  Fig.\ref{fig:spectrum}(a) shows the absorption spectrum of the $^1S_0\ |1/2, \pm 1/2 \rangle \leftrightarrow {^1}P_1\ \overline{|3/2, \pm 1/2 \rangle}$ transition detected using linearly polarized light along the $z$-axis. The blue line in Fig.\ref{fig:spectrum}(a) shows the absorption spectrum of the $^1S_0\ |1/2, \pm 1/2 \rangle\leftrightarrow {^1}P_1\ \overline{|3/2, \pm 1/2 \rangle}\ $ and $^1S_0\ |1/2, \pm 1/2 \rangle\leftrightarrow  {^1}P_1\ \overline{|3/2, \pm 3/2 \rangle}$ transition detected using linearly polarized light along the $y$-axis. In the presence of an electric field, due to the tensor Stark interaction, the degeneracy of the states $\overline{|3/2, \pm 1/2 \rangle}$ and $\overline{|3/2, \pm 3/2 \rangle}$ states of $6s6p\ ^1P_1$ is lifted, resulting in a double-peak structure in the absorption spectrum. By performing double-peak Lorentzian fitting, the centers of the transitions $^1S_0\ |1/2, \pm 1/2 \rangle \leftrightarrow {^1}P_1\ \overline{|3/2, \pm 1/2 \rangle}$ and $^1S_0\ |1/2, \pm 1/2 \rangle \leftrightarrow {^1}P_1\ \overline{|3/2, \pm 3/2 \rangle}$ can be obtained. Fig.\ref{fig:spectrum}(b) shows the absorption spectrum of the $^1S_0\ |1/2, \pm 1/2 \rangle \leftrightarrow {^1}P_1\ \overline{|1/2, \pm 1/2 \rangle}$ transition detected using linearly polarized light along the $y$-axis. The electric field is 19.96(3) kV/cm for the spectra measurement shown in Fig.\ref{fig:spectrum}.\par
First Section introduces the eigenenergies and eigenstates of atoms that change when interacting with an external electric field. Besides this, the E1 transition rate from the ground state $\ket{i}={}^1S_0\ \ket{1/2, 1/2}$ to an energy eigenstate $\ket{f}={}^1P_1\ \overline{\ket{F, m_F}}$ is also affected by the electric field. The expression for the transition rate is given by
\begin{equation}
      R_{i\xrightarrow{}f}\propto \left|\braketOP{i}{- \boldsymbol{d}  \cdot \boldsymbol{\epsilon} }{f}\right|^2=|D_{if}|^2
\end{equation}
where the E1 transition matrix element is
\begin{equation}\label{transition matrix}
\begin{aligned}
       D_{if}
    =\braketReOP{{}^1S_0}{\bm d}{{}^1P_1}\sum_{q}\sum_{F' m_F'}(-1)^{I+F'+q}\epsilon_{-q}\sqrt{2(2F'+1)}
            &\left(
            \begin{array}{ccc}
                1/2 & 1 & F '\\
                -1/2 & q & m_F' \\
            \end{array}
        \right)
        \left\{
            \begin{array}{ccc}
                1/2 & 1 & F '\\
                1 & 1/2 & 0 \\
            \end{array}
        \right\}
        \langle F', m_F'|\overline{F, m_F\rangle},
\end{aligned}
\end{equation}
where $\braketReOP{{}^1S_0}{\bm d}{{}^1P_1}$ is the reduced matrix element, and $\langle F',m_F'|\overline{F, m_F\rangle}$ is the projection of the energy eigenstate $\overline{|F', m_F'\rangle}$ onto the angular momentum eigenstate ${|F', m_F'\rangle}$. $\bm \epsilon$ is the unit vector of the polarization of the light electric field, $\bm d=-|e|\bm r$ is the electric dipole operator, and the subscript $q$ denotes the spherical tensor component. \par
\begin{figure}
\centering
\includegraphics[width=0.5\textwidth]{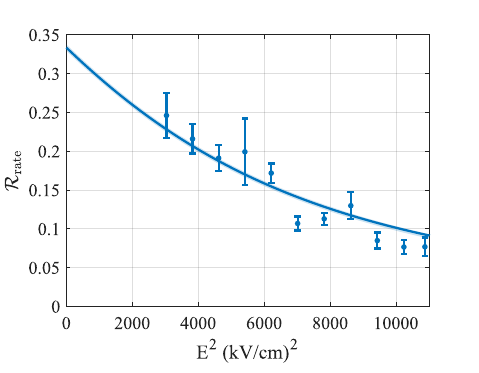}
\caption{$\mathcal{R}_{\rm rate}$ vs. $E^2$. The dependence of transition rates on the E field. $\mathcal{R}_{\rm rate}$ of the vertical axis is the ratio between the transition rates of $^1S_0|1/2,\pm1/2\rangle$ $\leftrightarrow$ $^1P_1|\overline{3/2, \mp1/2\rangle}$ and $^1S_0|1/2,\pm1/2\rangle$ $\leftrightarrow$ $^1P_1|3/2, \pm3/2\rangle$. The calculated results (solid line) agree with the data points. The error bars come from fitting to the signal shown in Fig. \ref{fig:spectrum}(a)}
\label{fig:transition rate} 
\end{figure}
\begin{figure}
\centering
\includegraphics[width=0.9\textwidth]{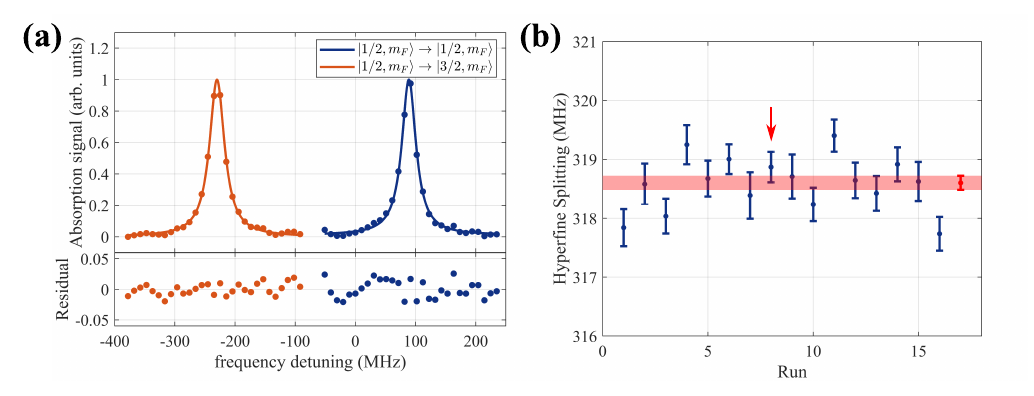}
\caption{Hyperfine structure constant measurement. (a) Spectra obtained in a single measurement. The orange line represents the transition \( 6s^2\, ^1\!S_0 |1/2, m_F \rangle \rightarrow 6s6p\, ^1\!P_1 |3/2, m_F \rangle \), with the fitted resonance and linewidth at -229.96(14) MHz and 29 MHz, respectively. The blue line represents the transition \( 6s^2\, ^1\!S_0 |1/2, m_F \rangle \rightarrow 6s6p\, ^1\!P_1 |1/2, m_F \rangle \), with the resonance and linewidth at 88.91(22) MHz and 29 MHz, respectively. The separation between the resonances is 318.87(26) MHz. (b) Results of 16 measurements. The red data point represents the weighted average of 16 measurements, yielding a value of 318.60(12) MHz, whose uncertainty is scaled by a factor of $\sqrt{\chi^2/\nu}=1.6$. The spectrum in Fig. S4(a) corresponds to the data point marked by the red arrow in S4(b).}
\label{fig:hyperfine}
\end{figure}

Fig.\ref{fig:transition rate} displays the ratio of transition rates ($\mathcal{R}_{\rm rate}$) between transitions $^1S_0|1/2,\pm1/2\rangle$ $\leftrightarrow$ $^1P_1|\overline{3/2, \mp1/2\rangle}$ and $^1S_0|1/2,\pm1/2\rangle$ $\leftrightarrow$ $^1P_1|3/2, \pm3/2\rangle$. The measured data agree with calculated $\mathcal{R}_{\rm rate}$ values (blue line).
Using the electric Breit-Rabi formula, the differential static scalar polarizability ($\Delta\alpha_s$) and the static tensor polarizability ($\alpha_{t}$) can be independently obtained from either the transition $^1S_0|1/2,\pm1/2\rangle$ $\leftrightarrow$ $^1P_1|\overline{3/2, \pm1/2\rangle}$ or $^1S_0|1/2,\pm1/2\rangle$ $\leftrightarrow$ $^1P_1|\overline{1/2, \mp1/2\rangle}$ as shown in the Table \ref{tab:table1}. Those value are consistent with each other, and agree with the values obtained on even isotopes by Kawamura $et\ al$.\cite{Kawamura2013}.
\begin{table}\centering
\caption{Polarizability values.}\label{tab:table1}
\begin{ruledtabular}
\begin{tabular}{ccc}
\textrm{Source} & \textrm{$\Delta\alpha_s$ (kHz/(kV/cm)$^2$}) & \textrm{$\alpha_t$ (kHz/(kV/cm)$^2$})  \\\hline

$^1S_0|1/2,\pm1/2\rangle$ $\leftrightarrow$ $^1P_1|\overline{3/2, \pm1/2\rangle}$
 
& $57.79\pm0.17_{\rm stat}\pm0.19_{\rm syst}$ & $-19.44\pm0.11_{\rm stat}\pm0.06_{\rm syst}$  \\

$^1S_0|1/2,\pm1/2\rangle$ $\leftrightarrow$ $^1P_1|\overline{1/2, \mp1/2\rangle}$
 
 & $58.03\pm0.08_{\rm stat}\pm0.19_{\rm syst}$ & $-19.22\pm0.13_{\rm stat}\pm0.06_{\rm syst}$ \\
Previous works\cite{Kawamura2013,Rinkleff1980} & 59.12(67)  & -19.47(17),-14.3(1.4) \\

\end{tabular}
\end{ruledtabular}
\end{table}
To determine the hyperfine structure constant \( A_{\rm hfs} \), we scan the absorption spectra of \( 6s^2\, ^1\!S_0 |1/2, m_F \rangle \rightarrow 6s6p\, ^1\!P_1 |1/2, m_F \rangle \) and \( 6s^2\, ^1\!S_0 |1/2, m_F \rangle \rightarrow 6s6p\, ^1\!P_1 |3/2, m_F \rangle \) with the electric field turned off (Fig.~\ref{fig:hyperfine}(a)), and measure the frequency difference between the two resonances. By performing repeated measurements, we obtain a series of hyperfine splitting results (Fig.~\ref{fig:hyperfine}(b)). Hence, the value of the hyperfine structure constant $A_{\rm hfs}$ is determined to be -212.4(1) MHz.

\vspace{\baselineskip}
\begin{acknowledgments}
We would like to thank W.-T. Luo, J.-L. Zhang, L.-Y. Tang, Z.-C. Yan, W. Jiang, C.-L. Zou for helpful discussions. This work has been supported by the National Natural Science Foundation of China (NSFC) through Grants No. 12174371, and the Innovation Program for Quantum Science and Technology through Grant No. 2021ZD0303101. 
\end{acknowledgments}

% $\textit{Acknowledgement\ -\ }$
\nocite{}
%%%%%%%%%% If using BibTeX:
\bibliography{electric_Breit_Rabi}

\end{document}